\documentclass[
  aip,
  APL,
  preprint,
  amsmath,amssymb,
  longbibliography
]{revtex4-2}

\usepackage{graphicx}
\usepackage{dcolumn}
\usepackage{bm}

\usepackage[separate-uncertainty=true]{siunitx} 
\usepackage{cleveref} 

\newcommand{\gev}{GeV$^-$}

\begin{document}

\title{Radiotracer photoluminescence for element-specific identification of color centers}

\author{Brecht Biesmans}
\affiliation{KU Leuven, Quantum Solid-State Physics, Celestijnenlaan 200D, 3001 Leuven, Belgium}

\author{Afonso Lamelas}
\affiliation{CICECO- Instituto de Materiais de Aveiro, Universidade de Aveiro, 3810-193 Aveiro, Portugal}

\author{Kirill Danilov}
\affiliation{KU Leuven, Quantum Solid-State Physics, Celestijnenlaan 200D, 3001 Leuven, Belgium}

\author{Aleksandr Seliverstov}
\affiliation{KU Leuven, Quantum Solid-State Physics, Celestijnenlaan 200D, 3001 Leuven, Belgium}

\author{\^Angelo Costa}
\affiliation{Centro de Ciências e Tecnologias Nucleares, Departamento de Engenharia e Ciências Nucleares, Instituto Superior Técnico, Universidade de Lisboa, Estrada Nacional 10, 2695-066 Bobadela, Portugal}

\author{V{\'i}tor Amaral}
\affiliation{CICECO- Instituto de Materiais de Aveiro, Universidade de Aveiro, 3810-193 Aveiro, Portugal}

\author{Andr{\'e} Vantomme}
\affiliation{KU Leuven, Quantum Solid-State Physics, Celestijnenlaan 200D, 3001 Leuven, Belgium}

\author{Jo{\~a}o Guilherme Correia}
\affiliation{Centro de Ciências e Tecnologias Nucleares, Departamento de Engenharia e Ciências Nucleares, Instituto Superior Técnico, Universidade de Lisboa, Estrada Nacional 10, 2695-066 Bobadela, Portugal}

\author{Ulrich Wahl}
\affiliation{Centro de Ciências e Tecnologias Nucleares, Departamento de Engenharia e Ciências Nucleares, Instituto Superior Técnico, Universidade de Lisboa, Estrada Nacional 10, 2695-066 Bobadela, Portugal}

\author{Lino M. C. Pereira}
\email{lino.pereira@kuleuven.be}
\affiliation{KU Leuven, Quantum Solid-State Physics, Celestijnenlaan 200D, 3001 Leuven, Belgium}

\author{the ISOLDE Collaboration}

\date{\today}

\begin{abstract}
We report on the implementation of a radiotracer photoluminescence spectroscopy setup at the ISOLDE radioactive ion beam facility at CERN, enabling element-specific identification of optically active defects in solids. The method combines radioactive ion implantation with optical spectroscopy, allowing the temporal evolution of photoluminescence signals to be correlated directly with nuclear decay. The setup is currently optimized for color centers in diamond and related wide-bandgap materials and enables room-temperature measurements. The system consists of an optical microscope coupled to a fiber-fed Czerny–Turner spectrometer with a liquid-nitrogen-cooled CCD detector, providing the stability required for long-duration measurements. As a proof-of-principle, radioactive $^{75}$Ga was implanted into diamond as a precursor to produce $^{75}$Ge impurities. The photoluminescence band extending from 600~nm, corresponding to the well-known GeV$^{-}$ center, exhibits an exponential decay with a half-life of \SI{82.3(2.5:2.3)}{\minute}, in agreement with the known $\beta^{-}$ decay half-life of $^{75}$Ge of \SI{82.78(4)}{\minute}. This establishes a direct and unambiguous correlation between the observed spectral feature and its germanium origin. These results demonstrate the capability of the setup to perform element-specific optical spectroscopy and extend radiotracer methods to color centers in wide-bandgap materials, taking advantage of the uniquely broad range of radioactive isotopes available at ISOLDE.
\end{abstract}

\maketitle

Point defects in diamond that give rise to optically active color centers have emerged as an important platform for quantum technologies, including quantum sensing, quantum communication, and photonic quantum information processing.\cite{Aharonovich2016,Atature2018} Among the various classes of defects studied in diamond, the group-IV vacancy centers such as the SiV, GeV, SnV and PbV defects have become promising candidates for scalable quantum photonics and spin–photon interfaces. These centers inherit inversion symmetry from their $D_{\text{3d}}$ point group, which strongly suppresses spectral diffusion and results in narrow optical transitions with favorable coherence properties.\cite{Rogers2014,Iwasaki2015,bradac2019quantum, Trusheim2020}

The fluorescent emission of such point defects is a key experimental observable, whose characteristics are identified through photoluminescence spectroscopy (PL). The PL spectra of color centers typically contain a sharp zero-phonon line (ZPL) associated with optical transitions between the defects' electronic energy levels, accompanied by phonon sidebands (PSB). In many cases, the observation of a new ZPL in diamond is interpreted as evidence for a previously unknown defect complex. However, assigning a specific ZPL to a particular impurity-related defect remains a significant challenge. Photoluminescence spectroscopy is not intrinsically element-specific, and therefore different defects can in principle produce similar spectral signatures. Density-functional theory (DFT) calculations are often used to guide defect identification, but their predictive accuracy for absolute ZPL energy values remains limited.\cite{Gali2019} Consequently, many fluorescence lines observed in diamond remain unassigned, and in other cases, the attribution to a specific impurity remains tentative. Examples include a number of unidentified color centers reported throughout the diamond spectroscopy literature,\cite{Zaitsev2001,green2022diamond} as well as more recently observed defects involving the heavy group-IV elements. Specifically, lead-related vacancy centers have attracted significant interest due to their potentially favorable spin-orbit and optical properties, but their detailed optical signatures and charge-state assignments remain under active investigation.\cite{Trusheim2019,Tchernij2018} Establishing a direct, unambiguous link between a fluorescence feature and the chemical identity of the defect is therefore of fundamental importance.

Radiotracer photoluminescence spectroscopy (rPL) overcomes this limitation by introducing elemental specificity into optical spectroscopy. In this approach, a \textit{radioactive} isotope of the impurity of interest is implanted. When the radioactive parent isotope decays, the impurity transmutes into a chemically different daughter element. As a result, any color center involving the parent atom is either destroyed or transformed upon decay, causing the corresponding PL signal to decay in time following the radioactive decay law. The reverse also holds: PL features associated with the daughter element may grow in time according to the same decay law. By monitoring the temporal evolution of the PL spectrum, spectral features that correlate with the radioactive decay can be directly associated with the implanted element. In this way, rPL establishes a direct connection between a given PL line and the chemical species forming the defect. Furthermore, features that exhibit no time dependence are not related to the radioactive isotope.

Radiotracer photoluminescence spectroscopy is part of a broader family of radioactive-ion-based methods developed for defect and impurity studies at on-line isotope-separator facilities such as ISOLDE-CERN.\cite{Deicher2002,Deicher2007} In particular, the concept of rPL was introduced in the mid-1990s as a method to chemically identify luminescent defect centers by exploiting the characteristic decay of radioactive isotopes implanted in a host semiconductor.\cite{magerle1995radioactive,daly1995chemical}  Subsequent work demonstrated the applicability of this approach to a range of impurity-semiconductor systems, including studies using various radioactive isotopes produced at ISOLDE in a wide range of semiconductors: $^{111}$In$\rightarrow^{111}$Cd, $^{191}$Pt$\rightarrow^{191}$Ir, $^{193}$Au$\rightarrow^{193}$Pt and $^{7}$Be$\rightarrow^{7}$Li in Si, $^{111}$In$\rightarrow^{111}$Cd and $^{71}$As$\rightarrow^{71}$Ge$\rightarrow^{71}$Ga in GaAs, $^{111}$Ag$\rightarrow^{111}$Cd and $^{197}$Hg$\rightarrow^{197}$Au in GaN, and $^{111}$Ag$\rightarrow^{111}$Cd in CdTe.\cite{henry2000} Since then, the method has been demonstrated in a series of experiments, particularly on ZnO. Johnston \textit{et al.} used rPL to identify donor-related impurity lines associated with Ga in ZnO.\cite{Johnston2006} Subsequent work demonstrated that implantation of radioactive parents and monitoring of the daughter-dependent signal growth could be used to correlate specific luminescence features with Ge,\cite{Johnston2011} and later with Sn and Sb.\cite{Cullen2013} These previous implementations of rPL at ISOLDE focused primarily on impurity-related luminescence in semiconductors, most notably donor-bound excitonic transitions. Despite the clear potential, the technique has so far not been applied to color centers, where the optical emission originates from deep defect states in insulators, and for which such element-specific identification would be of major importance. Additionally, whereas earlier rPL studies on shallow dopants in semiconductors typically benefited from low-temperature operation to stabilize bound-exciton luminescence, color centers in insulators often remain optically active at room temperature, such that rPL measurements can be performed without requiring cryogenic conditions.

In the present work, we report on the implementation of a new rPL setup at ISOLDE-CERN specifically designed for the investigation of color centers in diamond. The setup enables monitoring of photoluminescence at room temperature following the implantation of radioactive ions and the subsequent thermal annealing required for optical activation of defects. 
Although the method is generally applicable to a wide range of materials and optically active defects, we focus here on color centers in diamond as a model system. As a proof-of-principle, we demonstrate the technique using the well-known GeV$^{-}$ center, formed by implanting the radioactive precursor $^{75}$Ga, which undergoes nuclear transmutation into $^{75}$Ge and then into stable $^{75}$As through $\beta^{-}$ decay. We show that the temporal evolution of the PL signal directly correlates with the radioactive decay, confirming the element-specific capability of the rPL approach, as well as the effective performance of the implemented setup.

Following implantation, the samples are thermally annealed to optically activate the defects. The selected isotopes must decay through a process that changes the proton number, such that the parent atom transmutes into a different daughter element. Depending on whether the monitored defect contains the parent or the daughter element of the decay chain, the corresponding PL signal decreases or increases with time according to the radioactive decay law. Consequently, the half-life of the implanted isotope sets the timescale of the observed PL signal evolution.

\begin{figure}
   \centering
   \includegraphics[width=0.7\linewidth]{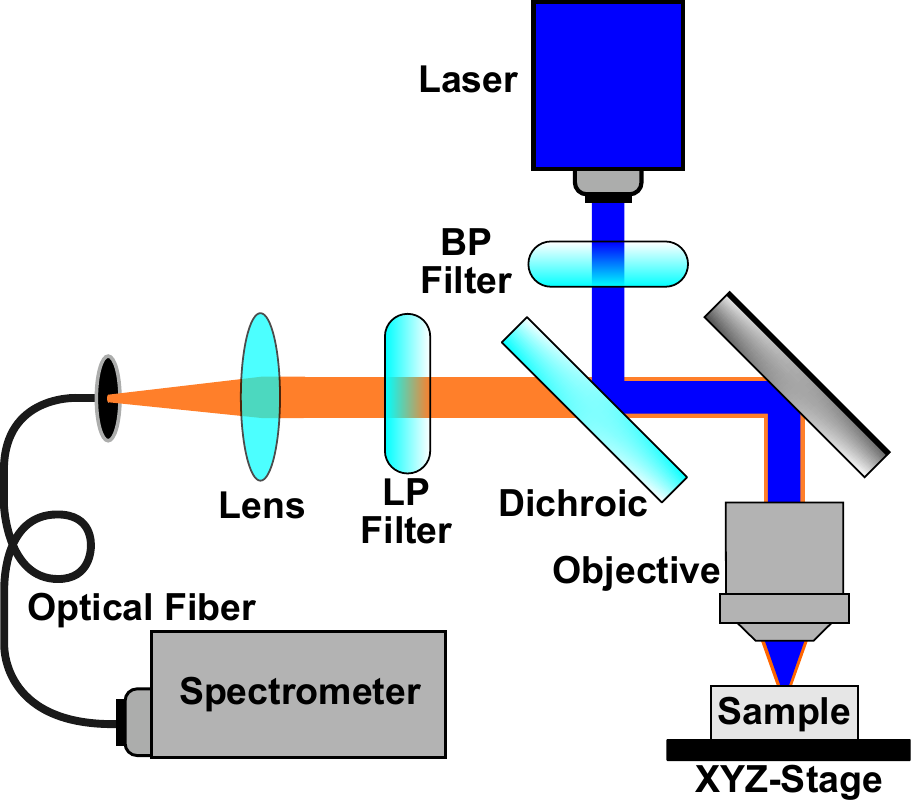}
   \caption{Schematic representation of the PL setup.}
   \label{fig:setup_schematic}
\end{figure}

The PL measurements are performed using a home-built setup as shown schematically in \cref{fig:setup_schematic}. A laser is first filtered with a narrow bandpass filter and then focused through a high numerical aperture microscope objective onto the diamond sample. Precise positioning is achieved by using an xy-stage for lateral movement and a z-stage for depth control. The fluorescence is collected using the same optical path, but filtered with a dichroic and longpass filter to separate excitation and emission light. Subsequently, the fluorescence is focused into an optical fiber with a doublet lens. Finally, the fluorescence is transmitted to the Czerny-Turner type spectrometer with an LN$_2$-cooled CCD for readout.

The rPL approach is illustrated here using the well-established photoluminescence of the GeV$^-$ center in diamond, characterized by a zero-phonon line (ZPL) at 602~nm and associated phonon sidebands.\cite{Iwasaki2015} The experiments were carried out at the ISOLDE facility at CERN.\cite{Catherall_2017} Since radioactive Ge beams are not available at ISOLDE, the short-lived precursor isotope $^{75}$Ga (half-life $t_{1/2} = 126(2) \mathrm{s}$) was implanted instead. It undergoes $\beta^-$ decay to $^{75}$Ge ($\beta^-$ endpoint energy 3300~keV, mean energy 1392~keV, nuclear recoil up to 102 eV), leading to the formation of GeV defects, as demonstrated previously.\cite{wahl2024structural} That study reported a GeV yield of approximately $20\%$ relative to the implanted particle concentration, with a reduction observed after thermal annealing. The probe isotope $^{75}$Ge ($t_{1/2} = 82.78(4)\mathrm{~min}$)\cite{NEGRET2013841} subsequently decays to stable $^{75}$As via $\beta^-$ emission (endpoint energy 1176~keV, mean energy 419.5~keV). The $^{75}$Ga beam was produced by bombarding a heated UC$_x$ target with 1.4~GeV protons, followed by out-diffusion, selective laser ionization of Ga,~\cite{Fedosseev_2017} and electromagnetic mass separation.

The experiment was performed on a $2 \times 2 \times 0.5$~mm$^3$ electronic-grade diamond (Element Six) with a nitrogen concentration of $[N] < 5$~ppb. The sample was implanted over 3~h at 30~keV to a fluence of $3 \times 10^{11}$~cm$^{-2}$ at an incidence angle of $10^\circ$ using a 2~mm diameter beam spot. This corresponds to an implantation depth of 149 \unit{\angstrom} with 36 \unit{\angstrom} of straggling, and a peak concentration of approximately $3 \times 10^{17}$~cm$^{-3}$. Following implantation, the sample was annealed at 900~$^\circ$C for 10~min in vacuum at a pressure below $5 \times 10^{-6}$~mbar. Given the 126~s half-life of $^{75}$Ga and the 50~min delay between the end of implantation and the start of annealing, the parent isotope had effectively fully decayed prior to the thermal treatment. The sample was then mounted in the PL setup for measurements, where preliminary surface and depth scans were performed to locate the optimal measurement coordinates. At the start of the time-series PL experiment, 1~h~25~min after the end of the implantation, 26\% of the $^{75}$Ge remained, corresponding to an effective fluence of $7.8 \times 10^{10}$~cm$^{-2}$.

\begin{figure}
    \centering
    \includegraphics[width=\linewidth]{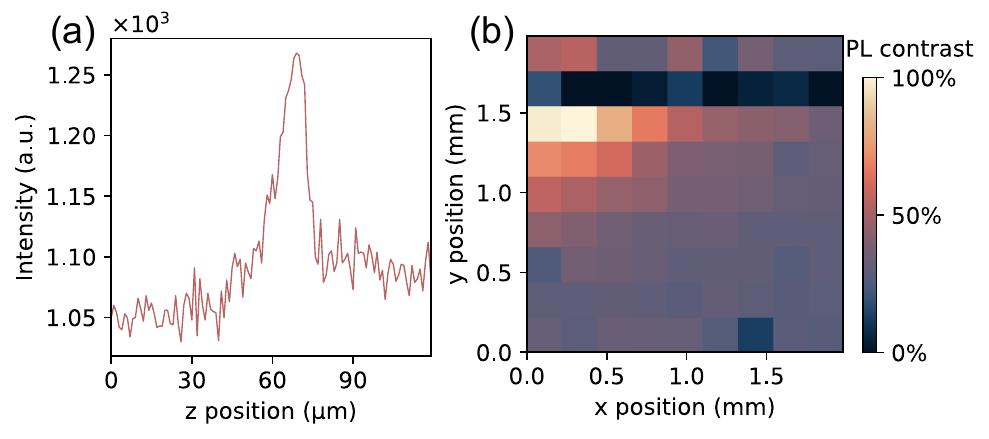}
    \caption{(a) PL intensity as a function of z position, with a maximum at 69 \unit{\micro\meter} and a full width at half maximum of $\sim$11~\unit{\micro\meter}. (b) PL contrast map showing the spatial distribution of the implanted region with a maximum at (0.248 \unit{\milli\meter}, 1.485 \unit{\milli\meter}). The color scale indicates the contrast of the GeV peak relative to the background. The horizontal black line arises from sample identification marks, which reduced the GeV fluorescence.}
    \label{fig:PL_map_rPL}
\end{figure}

PL measurements were performed at room temperature using continuous 457 nm excitation at 2~mW with the setup described in the supplementary material (section \ref{sec:75Ge_setup}). PL mapping ($xy$) was used to locate the implantation spot, followed by a depth ($z$) scan to identify the optimal position for subsequent measurements, both shown in \cref{fig:PL_map_rPL}. The depth dependence in \cref{fig:PL_map_rPL}(a) exhibits a surprisingly sharp maximum. This z‑sensitivity arises even though the setup is not confocal, and likely results from the low‑divergence 457 nm laser, focused by the high‑NA objective, that creates a near‑point‑like excitation volume. Moreover, the 200 µm core fiber may act as an effective pinhole that preferentially collects emission from that volume. 
The maximum of the curve effectively represents a focus very close under the surface and its width reflects the depth resolution of the setup, since the implantation profile is far narrower and shallower than the achieved depth resolution.
Accurate alignment is essential for the time-dependent measurements, which are performed in the region of highest PL intensity to maximize the observable duration of the decaying signal before reaching the detection threshold. Identifying the highest contrast spot in the $(x,y)$ PL map was particularly important, as the implantation profile was found to vary between implantations (examples of high-resolution scans on stable GeV exhibiting such variability are shown in \cref{fig:PL_map}). Time-series spectra were then acquired over approximately 8~h.


\begin{figure}
    \centering
    \includegraphics[width=\linewidth]{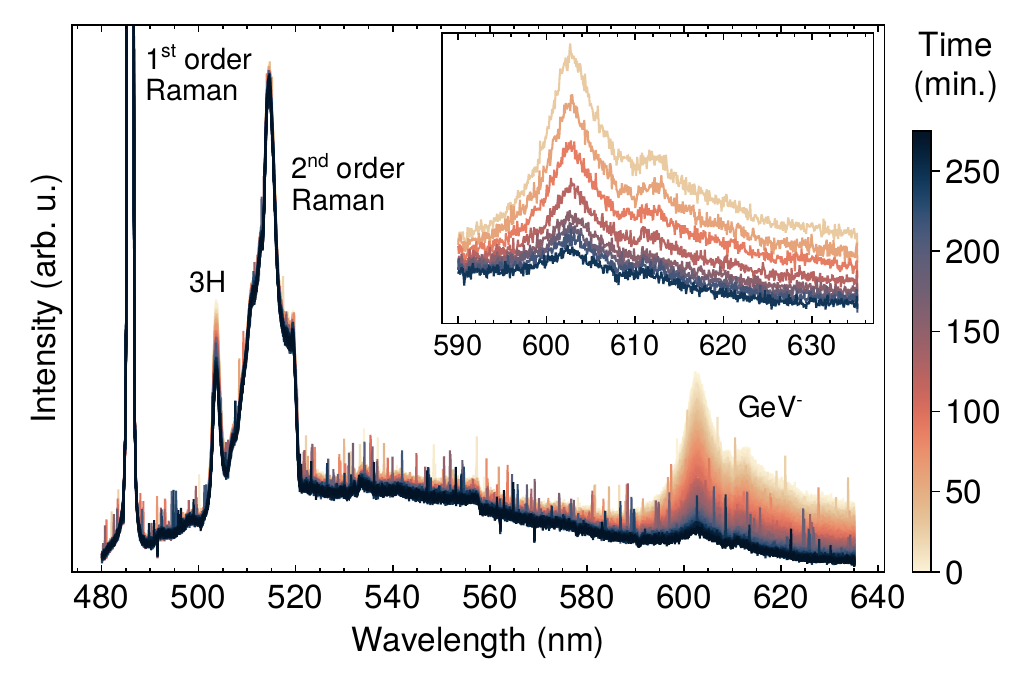}
    \caption{Measured spectra as a function of time with respect to the first measurement. The first (\SI{487}{\nm}) and second (\SI{507}{\nm}) order Raman signals from diamond are visible and are accompanied by a peak at \SI{503.6}{\nm} and the characteristic \gev~ZPL at \SI{602.8}{\nm}. The inset shows the latter, for a small subset of the time-steps for visual clarity.}
    \label{fig:spectra}
\end{figure}

\Cref{fig:spectra} shows the spectra measured as a function of the elapsed time since the start of the acquisition sequence.\cite{Zenodo75Ge} In addition to the characteristic first- and second-order Raman features of diamond, both the broad emission band and ZPL commonly associated with the \gev~center are observed from \SI{590}{\nm} onward and at \SI{602.8}{\nm}, respectively. An additional peak appears at about \SI{503.6}{\nm}, which we assign to the 3H center\cite{steeds_1999,vlasov_2002,pezzagna_2011}. This assignment is supported not only by the agreement with the reported ZPL wavelength, but also by the absence of a pronounced vibronic sideband at room temperature and the laser power-dependent quenching of the defect's fluorescence.

\begin{figure}
    \centering
    \includegraphics[width=\linewidth]{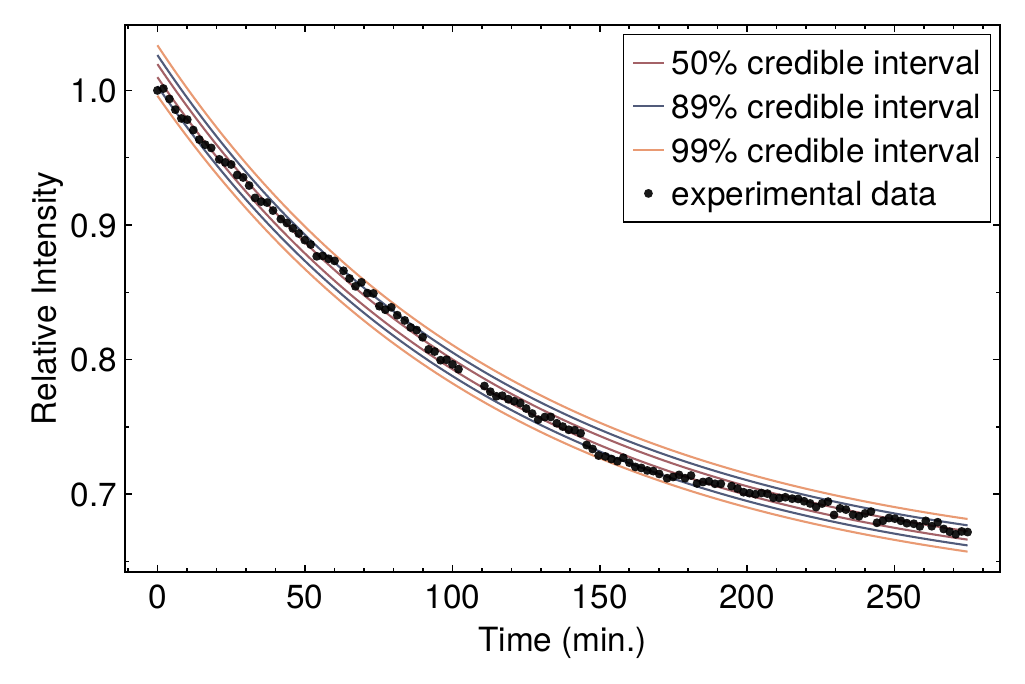}
    \caption{Summed intensity from the \SI{1}{\nano\meter} wide bin centered at \SI{602.8}{\nm} as a function of time normalized by the intensity of the first measurement. The 50\%, 89\% and 99\% equal-tailed credible intervals of the posterior predictive distribution are also shown.}
    \label{fig:intensity_decay}
\end{figure}

To model the decay of the signal intensity over time we included two components, one exponential, as expected from radioactive decay and another constant due to the background, which is composed of both dark counts from the detector as well as spectral signals that do not change in time. 
To extract the decay half-life, $t_{1/2}$, and fraction of the intensity that decays over time, $f$, we apply Bayesian inference because it provides robust parameter estimation with rigorous, self-contained uncertainty estimates, while naturally accommodating regions of the spectrum where no decay occurs without causing computational divergence due to unidentifiable parameters (more details in Supplemental Material).
To analyze the intensity decay as a function of wavelength, each experimental spectrum was individually binned into bins of \SI{1}{\nm} for which the spectral intensity was summed. An example is shown in \cref{fig:intensity_decay} for the bin centered at \SI{602.8}{\nm}. In the same figure, multiple equal-tailed credible intervals are represented which were estimated from the predictive posterior samples, indicating that the model used is able to explain the experimental data.

\begin{figure}
    \centering
    \includegraphics[width=\linewidth]{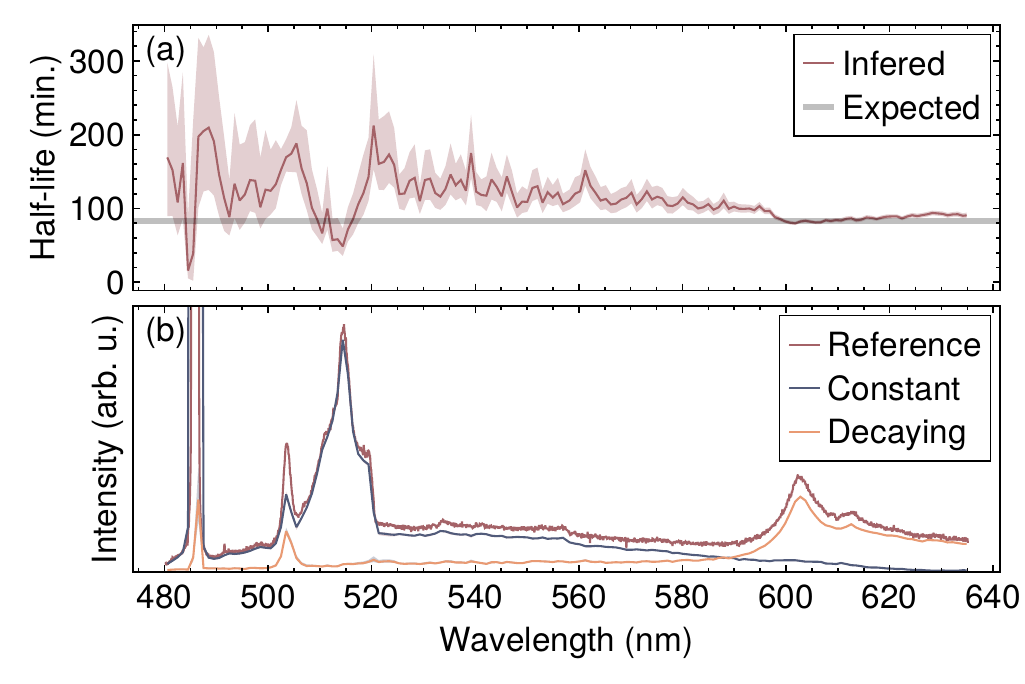}
    \caption{(a) Median posterior half-life ($t_{1/2}$) of the decaying signal and (b) photoluminescence signal decomposition into the component that is constant over time and the decaying one. In (a), the band represent the \SI{89}{\percent} equal-tailed credible interval of the respective variables and the horizontal line represents the half-life of $^{75}$Ge, \SI{82.78(4)}{\minute}. The decaying component in (b) was offset vertically for clarity and the first measured spectrum is also shown for reference of the spectral features.}
    \label{fig:spectral_decay}
\end{figure}

The parameters were estimated for each bin. The first (reference) measurement and the intensity attributed to its exponential and constant components are shown in \cref{fig:spectral_decay}, along with the estimated half-life. Starting from \SI{600}{\nm} almost all of the PL signal is decaying with a half-life in good agreement with the expected half-life from $^{75}\textrm{Ge}$, \SI{82.78(4)}{\minute}. For the peak at \SI{602.8}{\nm}, $t_{1/2}$ was estimated to be \SI{82.3(2.5:2.3)}{\minute}, where the interval corresponds to the \SI{89}{\percent} equal-tailed credible interval, which is in excellent agreement with the expected value. This result is insensitive to the choice of prior for the half-life (more details in Supplemental Material).
We therefore conclude that the emission band above \SI{600}{\nm} is associated with a germanium-related defect, thereby demonstrating the capability of rPL to link spectral features to the corresponding chemical element.
In the spectral region of both the first and second order Raman scattering, no significant decay is observed, so the fitted half-life is not meaningful. This is consistent with the expected constancy of the Raman signal and thus confirms the stability of the measurement system over time. The signal attributed to the 3H peak has a significant decay component with a half-life of \SI{170(25:20)}{\minute}.
Since the 3H center is well known in diamond samples without Ge,\cite{steeds_1999,vlasov_2002,pezzagna_2011} it cannot itself be considered Ge-related. Although 3H bleaching under blue illumination has been reported, the corresponding timescales are only seconds to minutes,\cite{steeds_1999,vlasov_2002} much shorter than the variation observed here. A possible explanation is that the presence of radioactive Ge in the diamond is affecting the natural bleaching of the 3H peak. This could occur through the disappearance of negatively charged GeV or the activity due to the electrons emitted from the $^{75}$Ge decay, which might affect the intrinsic charge dynamics of the 3H center.\cite{vlasov_2002} Additional considerations regarding the interpretation of rPL decay curves are provided in the Supplementary Material (section \ref{rPL note}).

In conclusion, we implemented a radiotracer photoluminescence (rPL) spectroscopy setup at ISOLDE-CERN that correlates the temporal evolution of photoluminescence from diamond samples containing radioactive impurities with the corresponding nuclear decay. As a proof of principle, implantation of $^{75}$Ga followed by decay to $^{75}$Ge and thermal annealing produced Ge-related color centers whose $\sim 600~\mathrm{nm}$ emission decayed with a half-life matching that of $^{75}$Ge, directly confirming its germanium origin and demonstrating element-specific PL assignment by rPL. These results validate both the performance of the newly implemented setup and the applicability of radiotracer photoluminescence spectroscopy to the study of vacancy–impurity color centers in diamond. More generally, the technique provides a powerful tool for the chemical identification of luminescent defects in wide-bandgap materials, which has often remained speculative. Future experiments exploiting other radioactive isotopes available at ISOLDE will allow rPL to be applied to a broad range of impurity-related color centers, offering a unique route to resolve long-standing questions regarding the microscopic origin of photoluminescence features in diamond and other materials. In particular, the approach can be extended to other impurity-related color centers in wide-bandgap materials, taking advantage of the uniquely broad range of radioactive isotopes available at ISOLDE. For example, group-IV vacancy centers in diamond and silicon carbide, rare-earth-related emitters, and transition-metal impurities are currently of significant interest for quantum technologies. 
\\


See the supplementary material for a description of the experimental setup, a discussion of the variability in implantation profiles, and more details of the Bayesian analysis framework used in this work. This section includes Refs.~\citenum{chopin_2020, zhou_2016, douc_2005, metropolis_1953, fearnhead_2013, talts_2020, modrak_2025}.
\\

We appreciate the support of the ISOLDE Collaboration and technical teams. This work was funded by the Fonds voor Wetenschappelijk Onderzoek, Vlaanderen (FWO, Flanders, G039624N, G089622N, I002619N), by the KU Leuven (IBOF/23/065), by the Portuguese Foundation for Science and Technology (FCT,
2024.00223.CERN (DOI 10.54499/2024.00223.CERN), UID/04349/2025 (DOI 10.54499/UID/04349/2025), BD/11398/2022, UIDB/50011/2020 (DOI 10.54499/
UIDB/50011/2020), LA/P/0006/2020 (DOI 10.54499/LA/P/0006/2020)). The EU Horizon Europe Framework supported ISOLDE beam times through Grant Agreement 101057511 (EURO-LABS).

\section*{AUTHOR DECLARATIONS}
\subsection*{Conflict of Interest}
The authors have no conflicts to disclose.

\subsection*{Author Contributions}
Brecht Biesmans and Afonso Lamelas contributed equally to this work.

\section*{Data Availability}
The data that support the findings of this study are available in Zenodo at \text{https://doi.org/10.5281/zenodo.20610439}.\cite{Zenodo75Ge} The repository includes the primary data and supplementary documentation used in this work. These materials are publicly accessible without restriction.

\bibliography{bib}

\clearpage
\onecolumngrid
\setcounter{page}{1}

\renewcommand{\thepage}{\arabic{page}}
\setcounter{section}{0}
\renewcommand{\thefigure}{S\arabic{figure}}
\setcounter{figure}{0}
\renewcommand{\thetable}{S\arabic{table}}
\setcounter{table}{0}
\renewcommand{\thesection}{S\arabic{section}}
\setcounter{section}{0}

\begin{center}
    \large\textbf{Supplementary Material for: \\ Radiotracer photoluminescence for element-specific identification of color centers}
\end{center}

\begin{center}

\textbf{Brecht Biesmans$^{1}$, Afonso Lamelas$^{2}$, Kirill Danilov$^{1}$, \^Angelo Costa$^{3}$, V\'itor Amaral$^{2}$, Andr\'e Vantomme$^{1}$, Jo\~ao Guilherme Correia$^{3}$, Ulrich Wahl$^{3}$, Lino M. C. Pereira$^{1,*}$, and the ISOLDE Collaboration}

\vspace{0.5em}

{\small
$^{1}$ KU Leuven, Quantum Solid-State Physics, Celestijnenlaan 200D, 3001 Leuven, Belgium \\
$^{2}$ CICECO- Instituto de Materiais de Aveiro, Universidade de Aveiro, 3810-193 Aveiro, Portugal \\
$^{3}$ Centro de Ciências e Tecnologias Nucleares, Departamento de Engenharia e Ciências Nucleares, Instituto Superior Técnico, Universidade de Lisboa, Estrada Nacional 10, 2695-066 Bobadela, Portugal \\
$^{*}$ Corresponding author: lino.pereira@kuleuven.be
}

\end{center}

\section{Photoluminescence Spectroscopy Setup} \label{sec:75Ge_setup}
For the present study, a standard PL microscopy configuration was chosen instead of a confocal implementation, since the measurements target ensemble color-center behavior rather than single-defect spectroscopy. The larger collection volume of a standard PL setup yields a higher total PL signal, which averages over small variations due to local inhomogeneities. Furthermore, standard PL systems are less sensitive to mechanical drift, vibrations, and long-term thermal fluctuations, making them better suited for extended measurements. The spatial resolution and out-of-focus fluorescence rejection are reduced compared to a confocal configuration. However, these trade-offs are acceptable as the long-term stability of the PL measurements is paramount for reliably tracking PL intensity changes over the timescales relevant to isotopic decay.

The photoluminescence measurements were performed using a custom-built PL
setup depicted in \cref{fig:setup_schematic} of the main text. The excitation source was a Cobolt 08-DPL 457 nm laser, filtered by a Thorlabs FBH 460-10 laserline filter to ensure spectral purity. The beam was directed onto the sample via a T470lpxr dichroic mirror and focused using an Olympus MPLAPON100X 0.95 NA AIR objective. The sample was mounted on an XYZ-stage comprising a motorized Thorlabs M30XY/M XY-stage and a piezo-driven Thorlabs PDXZ1/M Z-stage with PDXC2 controller for precise positioning. The emitted photoluminescence was collected through the same objective, filtered by the dichroic mirror and a Chroma ET460lp longpass filter, then transmitted via optical fiber (0.22 NA, 200 \unit{\micro\meter} core) into a Horiba IHR550 spectrometer equipped with a LN$_2$-cooled Horiba Spectrum ONE CCD3000 detector for high-resolution spectral analysis.

\section{Beam profile}

\begin{figure}[h]
    \centering
    \includegraphics[width=0.88\linewidth]{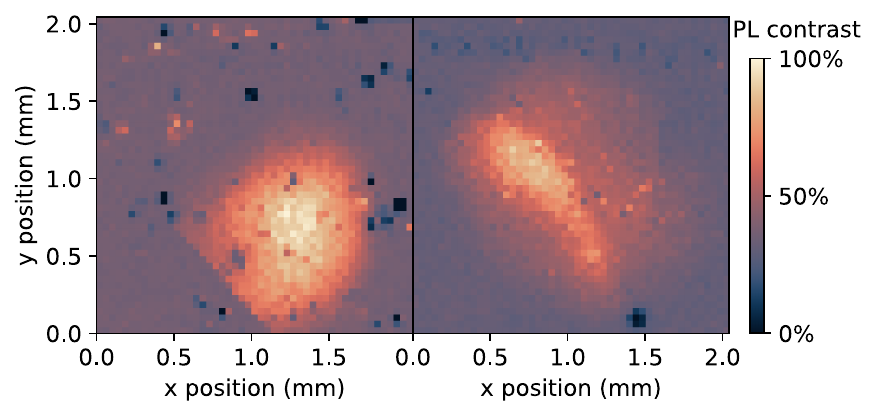}
    \caption{PL surface maps ($xy$) of two samples implanted using the same aperture of 1 mm diameter, but with different ion focusing parameters.}
    \label{fig:PL_map}
\end{figure}

\Cref{fig:PL_map} shows surface PL maps of two additional diamond samples, illustrating the typical variability in implantation spot profiles. Both samples were $2 \times 2 \times 0.5$~mm$^3$ electronic-grade diamonds (Element Six) with a nitrogen concentration $[N] < 5$~ppb, implanted with $^{74}$Ga ions ($t_{1/2} = 8.12(12)$~min), which undergo $\beta^-$ decay to $^{74}$Ge (endpoint energy 4777~keV, mean energy 990~keV, nuclear recoil up to 201 eV). The implantation was performed at 50~keV to a fluence of $1 \times 10^{11}$~cm$^{-2}$, at an incidence angle of $10^\circ$ with a 1~mm beam spot, corresponding to a simulated peak concentration of approximately $7 \times 10^{16}$~cm$^{-3}$. Once the implanted $^{74}$Ga had fully decayed, the samples were annealed at 900~$^\circ$C for 10~min in vacuum ($p < 5 \times 10^{-6}$~mbar). The differences in spatial profile between the two maps illustrate the variability due to fine ion beam focusing conditions, highlighting the importance of performing mapping and alignment before the rPL time-series measurements.

\section{Considerations for the Interpretation of rPL Decay Curves} \label{rPL note}
Inspection of the time‑series in \cref{fig:intensity_decay} of the main text shows that the decay curves exhibit slow modulations superimposed on the expected exponential behavior. These deviations are likely due to experimental instabilities (e.g., small drifts in laser power or z‑position) and to the intrinsic photophysics of the emitters. Photo‑bleaching or charge‑state conversion can cause a gradual loss or recovery of fluorescence that is unrelated to radioactive decay. More generally, rPL must be interpreted carefully: for instance, if the parent isotope recoils upon transmutation, it may create additional defects whose fluorescence masks the true decay signal. A well‑documented example is the green luminescence band in ZnO, which increases during $^{64}$Cu→$^{64}$Zn and $^{65}$Ni→$^{65}$Cu decay because the daughter nuclei recoil from substitutional Zn sites, likely creating Zn vacancies that emit in the green.\cite{Deicher2007} Careful control of laser stability, mechanical drift, and charge dynamics, along with complementary measurements, is therefore essential when interpreting rPL data.

\section{Bayesian Inference } 

To model the decay of the PL intensity as a function of time, we consider the sum of an exponential decay and a constant background. This framework allows us to extract the time-dependent signal decay even in the presence of time-independent contributions, such as detector dark counts or stable spectral features from other defects. For a significant portion of the spectrum, no temporal change is expected. In this limit ($f \rightarrow 0$), the decay half-life no longer influences the model, meaning a well-defined optimum cannot be resolved using a conventional least-squares approach. Furthermore, frequentist variance estimates at the final optimization point, which rely on the local Hessian matrix (second derivatives), tend toward infinity as the decay amplitude approaches zero. To overcome these limitations, we employ Bayesian inference. This approach naturally accommodates the stable limit by incorporating prior information, while providing robust parameter estimation and rigorous, self-contained uncertainty intervals without relying on asymptotic normality assumptions.

Fundamentally, Bayesian inference treats unknown parameters—such as the decay half-life ($t_{1/2}$) and the decaying fraction ($f$) as random variables defined by probability distributions rather than fixed, single-valued constants. The analysis begins by assigning a prior distribution to each parameter, which formally quantifies our initial knowledge or physical constraints before examining the data. By applying Bayes' theorem, these priors are systematically updated with the likelihood of the observed experimental data to yield the posterior distribution. This posterior distribution represents the complete, updated state of knowledge regarding our parameters, capturing not only their most probable values but also their dependency structures and uncertainties. The specific mathematical formulation of this probabilistic model is detailed below. 

The experimental data consist of pairs $(t_k, I_k)$ for each measured wavelength, where $t_k$ is the time elapsed since the first measurement, and $I_k$ is the recorded intensity normalized by the initial intensity, for $k$ from 1 to the number of measurements. Since the analysis is performed independently for each wavelength channel, we omit the wavelength index for notation brevity. We define the probabilistic model for the intensity decay as:  
\begin{equation}
    \begin{split}
        I_k &\sim \textrm{Gamma}\left( s^{-2}, m_k \cdot s^{2} \right)\\
        m_k &= I_0 \cdot \left( f \cdot \exp\left[-\ln{2} \cdot t_k / t_{1/2} \right] + (1-f)\right) \\
        I_0 &\sim \mathcal{T}[\text{Normal}(1,0.05),0,\infty] \\
        f &\sim \textrm{Beta}(0.5,5) \\
        t_{1/2} &\sim \mathcal{T}[\text{Normal}(0,120),0,\infty] \quad \text{(min.)} \\
        \ln(s) &\sim \text{Normal}(-5,1)
     \end{split}
     \label{eq:prob_model}
\end{equation}
The measured intensity $I_k$ is modeled using a Gamma distribution. Physically, the raw spectral counts collected by our CCD detector follow Poisson statistics (inherent to photon counting). Because we divide these raw counts by a constant to normalize the first measurement, the resulting data is no longer strictly integer-valued, but its statistical behavior is captured by a continuous Gamma distribution. The parameters are such that $m_k$ represents the expected mean intensity, and $s^2$ dictates the relative variance. All data points are treated as statistically independent.  $I_0$ represents the intensity at $t=0$, $f$ the fraction of that intensity that is due to the exponential decay and $t_{1/2}$ its half-life. The prior for $I_0$ is a Normal distribution centered around $1$ and with a small standard deviation due to the normalization by the first measurement. The prior for the decay fraction $f$ utilizes a Beta distribution heavily weighted toward $0$, reflecting our physical expectation that most of the overall diamond emission spectrum consists of a time-independent background rather than decaying radiotracer signals. The prior for $s$ is non-informative and selected to give reasonable prior predictive samples. The sensitivity of the results to the prior for $t_{1/2}$ is analyzed below.
$\mathcal{T}[\cdot,l,u]$ indicates that the corresponding distribution is truncated to the interval $[l,u]$ and the tilde in \cref{eq:prob_model} means distributed as.

The decay was parametrized in terms of $I_0$ and $f$ instead of the amplitude of the decay and a background constant as it results in a posterior density geometry that is easier to sample from. Moreover, it has a smooth behavior in the limit of no decay ($f\rightarrow0$), which is expected when performing inference on regions that are constant over time.  

The resulting posterior probability distribution cannot be derived analytically. Instead, we draw samples from it using a custom implementation of adaptive tempering Sequential Monte Carlo \cite{chopin_2020}. This method uses a sequence of distributions to interpolate from the simple prior to the complicated posterior, transporting a set of samples across this path. To build the sequence we used tempering, which introduces an inverse temperature-like parameter that controls the strength of the data, ranging from $0$, meaning no influence from the data, to $1$, the posterior distribution. The steps in between were determined adaptively based on the conditional effective sample size (ESS) \cite{zhou_2016}, choosing the next temperature such that this quantity was $0.8N$, where $N$ is the number of particles. Resampling was performed when the ESS reached $0.5N$, using residual resampling \cite{douc_2005}. A random-walk Metropolis-Hastings \cite{metropolis_1953} kernel was used to rejuvenate the particles with its covariance tuned using the empirical covariance of the particles from the previous temperature. The global scale of the proposal was adapted using the approach from \citeauthor{fearnhead_2013} \cite{fearnhead_2013}. The number of steps in the Markov chain was adapted at each temperature until the average square jump distance of the particles stabilized.

\begin{figure}
    \centering
    \includegraphics[width=\linewidth]{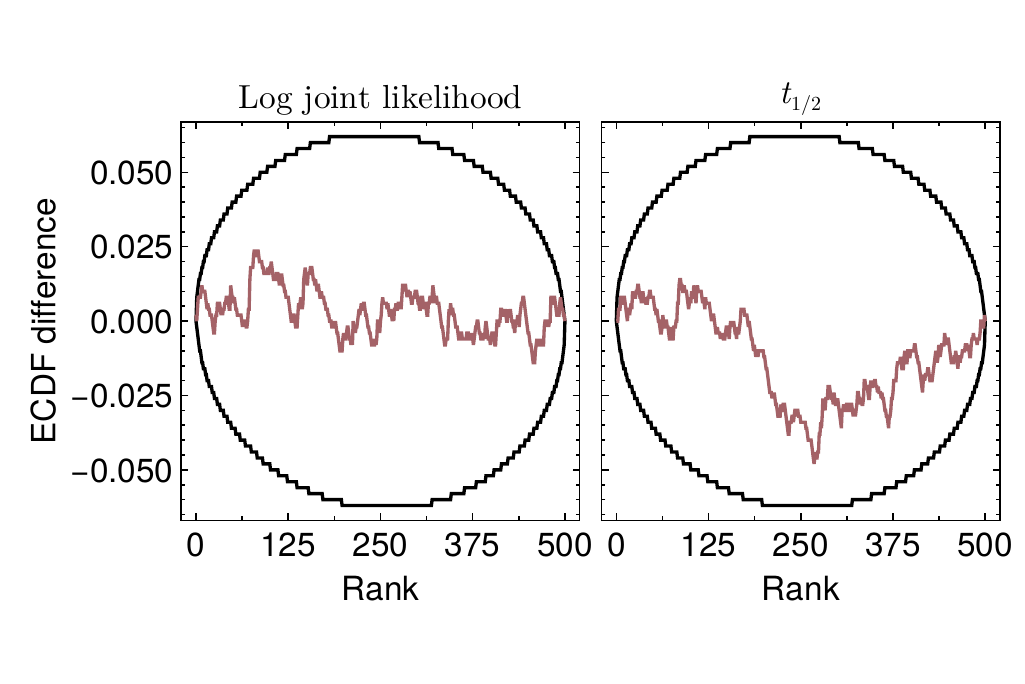}
    \caption{Computational validation of the Bayesian model using Simulation-Based Calibration (SBC). This diagnostic test acts as a check to verify that our inference procedure and model are mathematically consistent and unbiased. The plots show the deviation of our algorithm's performance from a theoretically perfect result, by the difference between the actual and expected empirical cumulative distribution function (ECDF) of the rank statistic. Two test quantities are represented, the log joint likelihood (a sensitive test quantity~\cite{modrak_2025}) and the half-life (which is the parameter of interest). Because the calculated data points stay entirely within the expected statistical boundaries (the black circles), which correspond to the \SI{89}{\percent} credible intervals, the test demonstrates that the inference procedure introduces no systematic errors and accurately extracts the physical parameters.}
    \label{fig:rank}
\end{figure}

To validate both the model and the inference procedure, simulation-based calibration (SBC) \cite{talts_2020,modrak_2025} was used. In SBC, we generate synthetic datasets using known parameters drawn from the prior and check if our Bayesian framework can accurately recover those starting values or derived quantities. By drawing $500$ simulated test cases and computing their posterior rank statistics, we generated the empirical cumulative distributions shown in \cref{fig:rank}. The results match the uniform distribution expected of an unbiased statistical tool, confirming that our model introduces no gross systematic errors and successfully recovers the physical parameters.  

\begin{figure}
    \includegraphics[width=\linewidth]{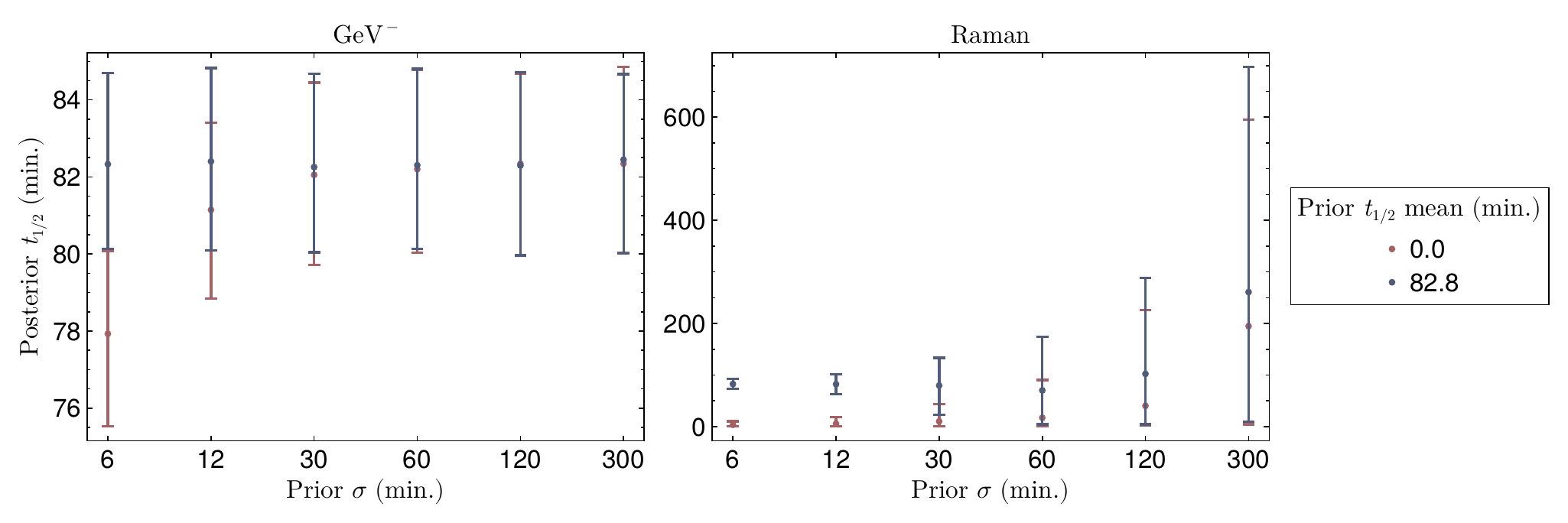}
    \caption{
    Sensitivity analysis of the estimated half-life ($t_{1/2}$) against changes in the prior distribution. This test evaluates whether our final physical conclusions are biased by the initial information. The posterior median and its accompanying $\SI{89}{\percent}$ credible interval are plotted as a function of the prior's mean and standard deviation for two distinct spectral regions: the $\textrm{GeV}^-$ peak ($\SI{602.8}{\nano\meter}$) and the Raman peak ($\SI{485.5}{\nano\meter}$).}
    \label{fig:prior_sensitivity}
\end{figure}

A common question in Bayesian analysis is whether the final answer is heavily biased by the initial information provided by the prior distribution. To test this robustness, we systematically varied the mean and standard deviation ($\sigma$) of the half-life prior across the entire spectrum. \Cref{fig:prior_sensitivity} highlights how our model reacts at the two primary regions of interest. 
We tested two extreme, representative scenarios from our data: the first-order Raman peak at $\SI{485.5}{\nano\meter}$ (where there is zero decay) and the peak of the $\textrm{GeV}^-$ signal at $\SI{602.8}{\nano\meter}$ (where the decay component is at its maximum).
At the $\textrm{GeV}^-$ peak ($\SI{602.8}{\nano\meter}$), if we force an extremely restrictive, narrow prior ($\sigma = 6$ or $12$ minutes), the final calculated half-life shifts slightly toward the prior choice (around \SI{5}{\percent}). However, as soon as the prior uncertainty is allowed to expand past $\sigma = \SI{30}{\minute}$, the final calculated half-life locks onto a stable value and becomes insensitive to how much wider we make the prior. This demonstrates that our experimental data is highly informative. The data completely overwhelms the prior, proving that the extracted half-life is a robust physical property determined by the experiment itself.  At the Raman Peak ($\SI{485.5}{\nano\meter}$), because the signal is physically constant, there is no decay behavior in the data. Consequently, the final uncertainty interval of the half-life is proportional to the width of the specified prior. This is the ideal and expected behavior when no decay exists ($f$ = \SI{0.0(0.4:0.0)}{\percent}), the data correctly informs the algorithm that a half-life parameter is physically meaningless.

\end{document}